\documentclass[12pt]{article}

\usepackage{psfig}
\usepackage{epsf}

\newcommand{\be}{\begin{eqnarray}}
\newcommand{\ee}{\end{eqnarray}}
\newcommand{\non}{\nonumber\\}
\newcommand{\bfu}{{\bf u}}
\newcommand{\bfv}{{\bf v}}
\newcommand{\bfw}{{\bf w}}
\newcommand{\bfx}{{\bf x}}
\newcommand{\bfy}{{\bf y}}
\newcommand{\bfz}{{\bf z}}
\newcommand{\calO}{{\cal O}}
\newcommand{\ave}[1]{\langle {#1} \rangle}
\newcommand{\bfQ}{{\bf Q}}

\newcommand{\bbox}[1]{\hbox{\boldmath{$#1$}}}

\title{
\vskip 0.1in
A classical Odderon in QCD at high energies}
\author{
Sangyong Jeon\\
{\small \it Physics Department, McGill University, Montreal, QC H3A-2T8,
Canada}\\
{\small \it  and RIKEN-BNL Research Center, Brookhaven National Laboratory,
Upton, NY 11973}
\\
Raju Venugopalan~\footnote{Present address: Institute for Theoretical Physics
II,
Univ. of Hamburg, Luruper Chaussee 49, 22607 Hamburg, Germany}\\
{\small \it Physics Department,
Brookhaven National Laboratory, Upton, NY 11973, USA.}\\
}

\begin{document}

\maketitle

\begin{center}
{\bf Abstract}\\
We show that the weight functional for color sources in the classical theory of
the Color Glass Condensate includes a term which generates Odderon excitations.
Remarkably, the classical origin
of these excitations can be traced to the random walk of partons in the two
dimensional space spanned by the SU(3) Casimirs.
This term is naturally suppressed for a large nucleus at high energies.

\end{center}

\noindent

\vfill \eject

\baselineskip=22pt plus 1pt minus 1pt
\parindent=25pt

\vskip 0.1in

\section{Introduction}
\vskip 0.1in

Hadronic cross-sections have been successfully described for some
time~\cite{DonnachieLandshoff} in terms of non-perturbative
t-channel exchanges known as
Pomerons and Reggeons. In perturbative QCD (pQCD), the Pomeron arises from the
color
singlet exchange of two reggeized gluons in the t-channel~\cite{BFKL}. This
BFKL Pomeron, leads to
cross-sections that grow rapidly as a power of the energy. Next-to-leading
order corrections to the BFKL Pomeron~\cite{FadinLipatov}, as well as
unitarization and saturation corrections, are topics of active interest in high
energy QCD~\cite{reviews}.

The Pomeron, which corresponds to a C-even exchange in the t-channel, has a
C-odd partner dubbed the
Odderon~\cite{Nicolescu}. In pQCD, it arises from the color
singlet exchange of three reggeized gluons in the t-channel~\cite{BKP}. This
BKP
Odderon, in contrast to the BFKL Pomeron, has solutions that grow at most
logarithmically as a function of the
energy~\cite{Odderon-solution}.  A nice review of the Odderon can be found in
Ref.~\cite{Carlo}.

Recently, Kovchegov et al. (KSW)~\cite{KSW}, working in the dipole
picture~\cite{Mueller,NZ},  computed non-linear saturation corrections to the
BKP
Odderon equation in the limit of  large number of colors $N_c$ and large atomic
number $A$. These are analogous to the saturation correction derived
previously~\cite{BK} for the BFKL Pomeron.  These
saturation effects modify the BKP result, leading  to an Odderon contribution
to high energy amplitudes that {\it decreases} with energy.

Understanding the
origins and role of the Odderon and its Pomeron partner, is essential for a
consistent formulation of high energy QCD. One such attempt
is the Reggeon effective field theory of Lipatov~\cite{Lipatov}. An alternative
approach is the Color Glass Condensate~\cite{IV} (CGC),
where the degrees of freedom are  parton fields at small x's of interest
coupled to
static parton sources at larger values of x~\cite{MV}; their  dynamical
evolution with x is described by the JIMWLK Wilsonian renormalization group
equations~\cite{JIMWLK}. Understanding the
correspondence between the two approaches will help clarify our understanding
of QCD at high energies.

With this aim in mind,  Hatta et al.~\cite{HIIM} identified the dipole and
baryon operators with Odderon
quantum numbers in the CGC framework. The JIMWLK equations for the evolution of
these operators are shown to be equivalent to the BKP equation
for low parton densities. In the large $N_c$ and large $A$ mean field
approximation,  the  KSW evolution equation for the Odderon dipole operator
is recovered. The JIMWLK equations,
in full generality, reveal a complex pattern of Pomeron-Odderon couplings under
quantum evolution. Ref.~\cite{HIIM} establishes a specific correspondence
between the framework of Reggeon field theory and the Color Glass Condensate.

The authors of Ref.~\cite{HIIM} also  argue that Odderon excitations are absent
in the McLerran-Venugopalan (MV) model~\cite{MV}, the classical effective
theory (without quantum evolution) of the CGC.
Their reasoning is correct for the original MV ansatz which assumed a Gaussian
random distribution of the large x color
sources in the CGC. However, we showed in a recent paper that the MV model for
an $SU(N_c)$ gauge theory where $N_c \geq 3$
includes additional contributions to the distribution of sources~\cite{SR}. In
this note, we will show that  these
contributions, sub-leading in the large $A$ limit, generate the Odderon.  For
SU(3), this term in the
path integral of the MV model is proportional to the cubic Casimir operator.
As in the small x limit,
where (quantum) saturation suppresses Odderon excitations, (classical)
saturation will be shown explicitly
to suppress these as well.

This letter is organized as follows. We begin by summarizing
briefly the MV model emphasizing
the coarse graining assumption implicit in the model.  We  state
recursion relations (derived previously in Ref.~\cite{SR}) which describe the
random
walk in color space of $SU(3)$ quark color sources. We extend the discussion of
Ref.~\cite{SR} to explicitly write down the path integral for the SU(3) MV
model.
With this path integral, we compute the Odderon operators identified in
Ref.~\cite{HIIM} both
in the weak field limit and to all orders in the classical field.
In the weak field limit, for $A=1$, our results for these agree with
Ref.~\cite{HIIM}.

\section{The MV path integral for a large nucleus and an SU(3) random walk}
\vskip 0.1in

The MV model is a simple model which captures several key features of QCD at
high parton densities~\cite{MV}. The model can be generalized and
the JIMWLK evolution equations we referred to previously arise from a proper
accounting of quantum corrections in the Leading Logaritmic Approximation in
x~\cite{JIMWLK}. The degrees of freedom are valence partons, which are
random, static, light cone sources, and  wee parton fields which couple
coherently to valence sources all along the Lorentz contracted
width of the nucleus. The kinematic condition for this coherence is
$x \ll A^{-1/3}$.
How many random sources the wee  parton fields couple to  depends on the
typical transverse momentum of the wee
parton~\footnote{The wee parton is soft only in its longitudinal
momentum-its transverse momentum may be large.}. A wee parton with
momentum $p_\perp$ resolves in the transverse plane of the nucleus, an area
$(\Delta x_\perp)^2 \sim 1/p_\perp^2$.
The number of valence partons it interacts simultaneously with is then
\be
k \equiv k_{(\Delta x_\perp)^2} = {N_{\rm valence}\over \pi R^2}\, (\Delta
x_\perp)^2
\, ,
\label{eq:1}
\ee
which indeed is proportional to $A^{1/3}$ since $N_{\rm
valence}=3\cdot A$ in QCD.  This counting of color charges is only
valid as long as $p_\perp > \Lambda_{\rm QCD}$.

The classical effective Lagrangean for the
MV-model, formulated in the infinite momentum frame ($P^+\rightarrow \infty$)
and light cone gauge ($A^+=0$) has the form,
\be
{\cal L} = S[A,\rho] + i W[\rho] \, .
\label{eq:2}
\ee
Here $S[A,\rho]$ has two terms. The first is the usual QCD Field Strength
tensor squared describing the
dynamics of the wee parton gauge fields $A^\mu$. The second describes the
coupling of these fields to the
source color charge density $\rho$. These have been discussed
previously~\cite{IV} and will not concern us further here.

Our focus will be on the final term-corresponding to a Boltzmann-like weight
$\exp( -W[\rho])$ in the generating functional for the
effective action. It denotes the likelihood of different $\rho$ configurations.
In the original MV model,
the weight functional for a very large nucleus had the Gaussian form
\be
W[\rho] = \int d^2 \bfx {\rho^a(\bfx)\rho^a (\bfx) \over 2\, \mu_A^2}
\, ,
\label{eq:3}
\ee
where the average color charge squared per unit area per color degree of
freedom, $\mu_A^2$ is simply determined to be~\cite{IV,SR},
\be
\mu_A^2 = {g^2 A\over 2\pi R^2} \,.
\label{eq:4}
\ee
It is of order $A^{1/3}$ fm$^{-2}$. For a large nucleus, this scale is the only
dimensionful scale in Eq.~\ref{eq:2}. Thus $\alpha_S(\mu_A^2) \ll 1$,
and all features of the theory can be computed in weak coupling.

In Ref.~\cite{SR}, it was shown explicitly that the assumption about the
Gaussian distribution of
charges for $k>> 1$ is exact only for quark sources in an SU(2) gauge
theory~\footnote{It is also true for gluon and quark--anti-quark sources in an
SU(3) gauge
theory. For a detailed discussion, see Ref.~\cite{SR}. This fact also explains
why only the Gaussian term was obtained in the computation of
Ref.~\cite{Kovchegov}, since there the nucleon was modeled as a
quark--anti-quark dipole.}. For SU(3) quarks, there is an additional
contribution to Eq.~\ref{eq:3}. Since the argument has
been stated at length previously in Ref.~\cite{SR}, we shall only outline key
features below with emphasis on novel results.

The problem of interest can be formulated as follows. Given $k$ random $SU(3)$
quarks in a coarse grained cell of transverse area $(\Delta x_\perp)^2$, what
is
the distribution of color representations? This problem can be formulated
straightforwardly as a recursion
problem.

When one adds a quark to an $(m,n)$ representation of SU(3),
the result is a sum of three irreducible
representations~\footnote{Representations in  $SU(3)$ can be expressed as
$(m,n)$, where $m$ and $n$ are
integers which label the upper and lower tensor indices respectively. A
fundamental quark ${\bf 3}$ state
in $SU(3)$ is  $(1,0)$ while an anti-quark in the ${\bf {\bar 3}}$ state
is $(0,1)$.} ,
 \be
 \bbox{
 (1, 0) \times (m, n)
 =
 (m+1, n) + (m-1, n+1) + (m, n-1)
 }
 \label{eq:5}
 \ee

One can deduce therefore that the multiplicity of a
particular representation $N_{m,n}^{(k+1)}$ in the
distribution of $k+1$ quarks (among all possible representations) is  given by
the recursion relation
 \be
 N^{(k+1)}_{m,n} =
 N^{(k)}_{m-1,n}
 +
 N^{(k)}_{m+1,n-1}
 +
 N^{(k)}_{m,n+1}
 \label{eq:6}
 \ee
 for $m, n \ge 0$ with the boundary condition
 $N^{(k)}_{m,-1} = N^{(k)}_{-1,n}=0$ and the initial condition,
 $N^{(0)}_{0,0} = 1$ {\it else} $N^{(0)}_{m,n} =0$.

As shown in Ref.\cite{SR},
the multiplicity $N_{m,n}^{(k)}$ can be expressed in terms of the tri-nomial
function $G_{k;m,n}$ defined as
 \be
G_{k:\; m, n} =
 {k!\over
 \left(k+2m+n\over 3\right)!
 \left(k-m+n\over 3\right)!
 \left(k-m-2n\over 3\right)!} \, .
 \label{eq:7}
 \ee
One finds that
 \be
 N_{m,n}^{(k)}
 & = &
 G_{k:\; m, n} + G_{k:\; m+3, n} + G_{k:\; m, n+3}
 \non
 & & {}
 -
 G_{k:\; m+2, n-1} - G_{k:\; m-1, n+2} - G_{k:\; m+2, n+2}
 \label{eq:8}
 \ee
satisfies the recursion relation Eq.~(\ref{eq:6}) with the stated initial and
boundary conditions.

The two Casimirs for a given $SU(3)$ $m, n$ state are the quadratic and cubic
Casimirs~\cite{Macfarlane},
 \be
 D_2^{(m,n)} &=&
 {1\over 3}\left(m^2 + mn + n^2\right) + (m + n) \nonumber \\
D_3^{(m,n)} &=& {1\over 18}
\left( m + 2 n + 3 \right)
\left( n + 2 m + 3 \right)\left(m - n\right) \, .
\label{eq:9}
\ee
Note that while the former is symmetric in $m$ and $n$, the latter is
anti-symmetric under exchange of $m$ and $n$.
For large $k$, one can use the Stirling formula to simplify Eq.~\ref{eq:7}, and
 further use Eq.~\ref{eq:9} to obtain the
very simple result:
 \be
 G_{k:\, m, n}
 & \approx &
 {\frac{{3^{{\frac{3}{2}} + k}}}{2\,k\,\pi }}
 \, \exp\left(-3 D_2^{m,n}/k \right)
 \left( 1 + 3 D_3^{m,n}/k^2\right)\, .
 \label{eq:10}
 \ee

The multiplicity, defined in terms of the $G$'s in Eq.~\ref{eq:8}, can be
approximated in the large $k$ limit as
\be
 N_{m,n}^{(k)}
 & \approx &
 2\, \partial_m^3 G_{k:\, m, n}
 +
 2\, \partial_n^3 G_{k:\, m, n}
 -
 3\, \partial_m \partial_n^2 G_{k:\, m, n}
 -
 3\, \partial_n \partial_m^2 G_{k:\, m, n}
 \non
 & \approx &
 {27\,m\,n\,\left( m + n \right) \over k^3}\,
 {\frac{{3^{{\frac{3}{2}} + k}}}{2\,k\,\pi }}
\, \exp\left(-3\, D_2^{m,n}/k \right)\,
 \left( 1 + 3 D_3^{m,n}/k^2\right)\, .
 \label{eq:11}
 \ee
 While Eq.~\ref{eq:10} was derived in Ref.~\cite{SR}, the final expression in
Eq.~\ref{eq:11} is new.
 The fact that the correction due to the cubic Casimir in Eq.~\ref{eq:11} is
identical to that in Eq.~\ref{eq:10}
 is non-trivial and remarkable. From Eq.~\ref{eq:9}, one can deduce that the
peak of the distribution is at $m = O(\sqrt{k})$ and $n =
O(\sqrt{k})$.  Correspondingly, the cubic Casimir introduces a correction of
size $O(1/\sqrt{k})$ for large k.

The dimension of an SU(3) representation is
 \be
 d_{mn} = {(m+1)(n+1)(m+n+2)\over 2} \approx {mn(m+n)\over 2}\, .
\label{eq:12}
 \ee
The probability to find the
system of $k$ quarks in an $(m,n)$ representation is therefore
proportional to $d_{mn}\, N^{(k)}_{m,n}$. One can then
define a normalized distribution~\footnote{One should
not interpret the integrand as a probability density when $m$
and $n$ are large compared to $\sqrt{k}$. But this is academic since the
probability
to find such large $m$'s and $n$'s  is exponentially suppressed.} in the sense
that
\be
1 = {27\sqrt{3}\over 4 k^4 \pi}
\int_0^\infty dm dn\, m^2 n^2 (m+n)\,
e^{-3\, D_2^{m,n}/k}\,
 \left( 1 + 3 D_3^{m,n}/k^2\right)\, .
\label{eq:13}
\ee
Since the cubic Casimir changes sign if $m$ and $n$ are exchanged, the
correction term gives a vanishing contribution to the integral.

In Ref.~\cite{SR}, we proved that the right hand side of Eq.~\ref{eq:13} can be
re-written {\it exactly} as an integral over classical color charges~\footnote{
Our derivation did not include the cubic Casimir but
that does not change anything since that term integrates to zero.},\be
 1 \approx \left(N_c\over k\pi \right)^{4}\int d^8 \bfQ\,
 e^{-N_c\, {\bfQ}^2/k + 3\,D_3({\bfQ})/k^2} \, ,
 \label{eq:14}
 \ee
where the cubic Casimir term in the integral is interpreted as a perturbation
to the dominant
quadratic Casimir term.  Here ${\bfQ} = (Q_1, Q_2, \cdots, Q_8)$
 is a classical color charge vector defined by
 $|{\bfQ}|=\sqrt{Q^aQ^a} \equiv \sqrt{D_2^{m,n}}$
 and $Q_1,\cdots,Q_8$ are its eight components.
 The cubic Casimir, in terms of the color charges,  is
$D_3(Q) = d_{abc} Q^a Q^b Q^c$,
 where $d_{abc}$ is the symmetric tensor in the $SU(3)$ Lie algebra.

$Q^a$ is the classical color charge in a single coarse grained box. One can
define a local classical
color charge density $\rho^a$ in terms of $Q^a$:
\be
Q_a = {1\over g}\int_{(\Delta x_\perp)^2} d^2 \bfx\,
\rho_a(\bfx)
\approx
{1\over g}(\Delta x_\perp)^2\, \rho_a(\bfx) \, .
\label{eq:16}
\ee
With this definition, and $k$ defined as in Eq.~\ref{eq:1},
one can write down a path integral for the distribution of the
color charges as
\be
1 \approx \int [d\rho]\,
\exp\left( -\int_A d^2 \bfx \left[{\rho_a(\bfx) \rho_a(\bfx)\over 2\mu_A^2}
-
{d_{abc}\, \rho^a(\bfx) \rho^b(\bfx) \rho^c(\bfx) \over
\kappa_A}\right]
\right) \, .
\label{eq:17}
\ee
where $\mu_A^2= g^2 A/ 2\pi R_A^2$ was defined in Eq.~\ref{eq:4}.  For the
weight of the cubic
Casimir, we obtain,
\be
\kappa_A = {g^3 A^2 N_c \over \pi^2 R^4 } \, .
\label{eq:18}
\ee
As we will now see, the cubic term in Eq.~\ref{eq:17} is precisely the term
that gives rise to the Odderon.

\section{Odderon in the classical effective theory}

In Ref.~\cite{HIIM}, the authors identified the Odderon exchange operators in
two situations, a) a dipole
scattering off the CGC, and b) a three quark baryon system scattering off the
CGC. We first consider the former where the C-odd ``dipole odderon operator" is
defined to be
\be
{\cal O}({\bf x}, {\bf y}) = {1\over 2i N_c}\,{\rm Tr}\left(V_x^\dagger V_y -
V_y^\dagger V_x\right) \, ,
\label{eq:19}
\ee
where
\be
V_x^\dagger \equiv V_{-\infty,\infty}^\dagger(\bfx)
=P\exp\left(ig\int_{-\infty}^\infty dx^- \alpha^a(x^-,\bfx) t^a\right) \, ,
\label{eq:20}
\ee
and $V_{-\infty,\infty}^\dagger = V_{\infty,-\infty}$.
Here $t^a$ is a generator in the fundamental representation of $SU(3)$ and
$\alpha^a$ is the CGC gauge field. In the weak field
limit, one can expand out the path ordered exponentials-the lowest order
contributing to the Odderon is  of cubic
order in $\alpha$~\cite{HIIM}:
\be
{\cal O}(\bfx,\bfy)\approx {-g^3 \over 24 N_c} d^{abc} (\alpha_x^a
-\alpha_y^a)\,(\alpha_x^b-\alpha_y^b)\,(\alpha_x^c-\alpha_y^c) \, .
\label{eq:21}
\ee
Now, from the classical equations of motion for gauge fields in the CGC, the
gauge field can be expressed in term of
the classical color charge density $\rho^a$ in Lorentz gauge as
\be
\alpha_x^a ={1\over 4\pi}\, \int d^2 {\bf z}\, \ln\left( {1\over ({\bf x}-{\bf
z})^2 \Lambda^2}\right)\, \rho^a({\bf z}) \, ,
\label{eq:22}
\ee
where $\Lambda$ is an infra-red cut-off which will not appear in the final
result.
 From this relation therefore, one expects ${\cal O}\propto
d^{abc}\rho^a\rho^b\rho^c$. In the original MV
ansatz, the expectation value of this Odderon operator would be zero.

However, from Eq.~\ref{eq:17}, we now know that there is indeed a
contribution (albeit suppressed) in the source
term of the MV model which has the color structure of the cubic Casimir
operator $d^{abc}\rho^a\rho^b\rho^c$.  One can therefore compute the initial
condition for
quantum evolution of the CGC directly from the MV model.

To evaluate $\ave{\calO}$, with the measure in Eq.~\ref{eq:17}, one needs to
compute the integral
\be
{\cal I}
&=& {d_{abc} d_{\bar{a}\bar{b}\bar{c}}\over Z_0 \,\kappa_A}
\int [d\rho]\, e^{-\int d^2\bfx {\rho^2/2\mu_A^2}}\non
&\times&
\int d^2 \bfy \,
\rho^a(\bfy)\rho^b(\bfy)\rho^c(\bfy)
\rho^{\bar{a}}(\bfu)\rho^{\bar{b}}(\bfv)\rho^{\bar{c}}(\bfw) \, .
\label{eq:23}
\ee
As prescribed previously, we have expanded out the cubic piece in
Eq.~\ref{eq:17} here. $Z_0$ corresponds to this measure.
${\cal I}$ can be easily computed using Wick's theorem.
Ignoring disconnected pieces for the moment, one obtains,
\be
{\cal I}
=
6\, {d_{abc} d_{\bar{a}\bar{b}\bar{c}}\over Z_0 \,\kappa_A}
\int d^2 \bfy \,
\ave{\rho^a(\bfy)\rho^{\bar{a}}(\bfu)}
\ave{\rho^b(\bfy)\rho^{\bar{b}}(\bfv)}
\ave{\rho^c(\bfy)\rho^{\bar{c}}(\bfw)}
\label{eq:24}
\ee
where we used the fact that $d_{abc}$ is a totally symmetric tensor. In the MV
model, the 2--point
correlator is simply
\be
\ave{\rho^a(\bfx)\rho^b(\bfy)} =
\mu_A^2\,\delta^{ab}\,\delta^{(2)}(\bfx - \bfy) \, .
\label{eq:25}
\ee

Using Eqs.~\ref{eq:22}-\ref{eq:25} to compute the expectation value of ${\cal
O}$, one obtains,
\be
\ave{\calO(\bfx,\bfy)}
&=&
{-g^3\,\mu_A^6\over 4\, N_c\,\kappa_A} d^{abc}\,d_{abc}
\int d^2\bfu
\left( G(\bfx-\bfu) - G(\bfy-\bfu) \right)^3
\non
\non
& = &
\alpha_S^3\,
{(N_c^2-4)(N_c^2-1)\over 4\,\pi\,r_0^2\,N_c^3}
A^{1/3}\,
\int {d^2\bfu}
\ln^3
{
\left| \bfx - \bfu \right|
\over
\left| \bfy - \bfu \right|
}\, ,
\label{eq:26}
\ee
where we defined $G(\bfx - \bfy) =
\ln\left(1\over (\bfx-\bfy)^2 \Lambda^2\right)/4\pi$. Also, note
that $\mu_A^2$ and $\kappa_A$ were defined in
Eq.~\ref{eq:4} and Eq.~\ref{eq:18} respectively, $r_0 = 1.12$ fm  and we used
the identity $d_{abc}\,d^{abc} = (N_c^2-1)\,(N_c^2-4)/N_c$.

In performing this computation,
we ignored the disconnected pieces coming from the Wick
expansion of Eq.(\ref{eq:23}). These pieces vanish because they involve a
summation with two contracted indices of $d_{abc}$.

We now turn to the Odderon operator corresponding to a three quark baryon state
scattering
off the CGC. In that case, the relevant
operator is~\cite{DoschCarlo,HIIM}
\be
{\cal B}({\bf x},{\bf y},{\bf z}) =
{1\over {3! \,2i}}\left({\epsilon}^{ijk}\,
{\epsilon}^{lmn}\,V_{il}^\dagger({\bf x})\,V_{jm}^\dagger({\bf y})\,
V_{kn}^\dagger({\bf z}) -h.c.\right)
\, .
\label{eq:27}
\ee
where $V_x^\dagger$ is defined in Eq.~\ref{eq:20}. Again, expanding out to
lowest non-trivial order in $\alpha$,
the expression for {\cal B}({\bf x},{\bf y},{\bf z})in the weak field limit is
\be
\approx {g^3\over
144}\,d^{abc}\left\{(\alpha_x^a +\alpha_y^a -2\alpha_z^a)\,(\alpha_y^b +
\alpha_z^b-2\alpha_x^b)\,(\alpha_z^c +\alpha_x^c-2\alpha_y^c)\right\} \, .
\label{eq:28}
\ee
Replacing $\alpha\rightarrow \rho$ using Eq.~\ref{eq:22}, using Wick's theorem
and computing
the average over color sources using Eq.~\ref{eq:25}, one obtains,
\be
\ave{\cal B}
& = &
{g^3\mu_A^6\over 144} d^{abc} d_{abc} {6\over \kappa_A}
\int d^2 \bfu\,
\left(G(\bfx-\bfu) + G(\bfy-\bfu) - 2
\,G(\bfz-\bfu)\right)\non
&\times&
\left(G(\bfz-\bfu) + G(\bfx-\bfu) - 2\,
G(\bfy-\bfu)\right)\non
&\times&
\left(G(\bfy-\bfu) + G(\bfz-\bfu) - 2\, G(\bfx-\bfu)\right)
\non
& = &
{\alpha_S^3 \over 24\,\pi\,r_0^2}
{(N_c^2-4)(N_c^2-1)\over N_c^2}\, A^{1/3}\,\int {d^2\bfu}\,
\ln{ |\bfx-\bfu||\bfy-\bfu|\over |\bfz-\bfu|^2}\non
&\times&
\ln{ |\bfy-\bfu||\bfz-\bfu|\over |\bfx-\bfu|^2}
\ln{ |\bfz-\bfu||\bfx-\bfu|\over |\bfy-\bfu|^2}\, .
\label{eq:29}
\ee
For $N_c=3$ and $A=1$, both Eq.~\ref{eq:26} and Eq.~\ref{eq:29}  can be shown
to agree exactly with Eqs.~5.5 and 6.2 respectively
in Ref.~\cite{HIIM}. To compare, we need to assume the nucleon is dilute and
the fact that the result of
Ref.~\cite{HIIM} is for a single quark (namely divide our result by 3).
In Eq.~\ref{eq:26} and Eq.~\ref{eq:29},  the factor of $A^{1/3}$
accounts for the possibility that the Odderon couples to $A^{1/3}$ nucleons in
the nucleus.

\section{Result to all orders in the gluon density}
We shall now explicitly evaluate the expectation value of the
dipole odderon operator in Eq.~\ref{eq:19}
to all orders in $g^2 \mu_A^2/k_\perp^2$.
In performing this computation, we have to take into account the path ordering
of the gauge field in the $x^-$ direction-see Eq.~\ref{eq:20}. The expectation
value of $\calO(\bfx, \bfy)$
is given by
\be
\left<{\cal O}\right>
 &=& {d_{abc} \over 2iN_c \kappa_A}
\int du^-
\int d^2 \bfu\,
\Big<
\rho_a(u^-, \bfu)
\rho_b(u^-, \bfu)
\rho_b(u^-, \bfu)\non
&\times&
{\rm Tr}\,\left(V(\bfx)^\dagger V(\bfy) -
V(\bfy)^\dagger V(\bfx)\right) \Big>
\label{eq:30}
\ee
where the average is now done with the ``smeared" Gaussian weight
\be
W[\rho]
=
\exp\left(-\int dz^-\int d^2 \bfz\,
{\rho_a(z^-, \bfz)\rho_a(z^-,\bfz)\over 2\,\mu^2(z^-)}
\right)
\label{eq:31}
\ee
Note that $\left<\rho^a(x^-,\bfx)\rho^b(y^-,\bfy)\right> =
\mu^2(x^-)\,\delta^{ab}\,\delta(x^--y^-)\,\delta^{(2)}(\bfx-\bfy)$.
To evaluate $\left<{\cal O}\right>$, we use the following identity,
\be
d_{abc}
{\delta \over \delta \rho_a(u)}
{\delta \over \delta \rho_b(u)}
{\delta \over \delta \rho_c(u)} W[\rho]
=
-{d_{abc}\over \mu^6(u^-)}
\rho_a(u)\rho_b(u)\rho_c(u)\,  W[\rho]
\label{eq:32}
\ee
using the fact that $W$ is a Gaussian and $d_{aab}$, summed over $a$ is zero.
Replacing the
LHS of this identity, which appears in Eq.~\ref{eq:30}, with the RHS and
integrating by parts, we can write
${\cal O}$ as the sum of two terms, where the first is
\be
\left<{\cal O}\right>_1
& = &
{d_{abc} \over 2i\,N_c\, \kappa_A}
\int du^-\, \mu^6(u^-)
\int d^2 \bfu\,
\Big<
{\delta \over \delta \rho_a(u)}
{\delta \over \delta \rho_b(u)}
{\delta \over \delta \rho_c(u)}\non
&\times&
{\rm Tr}\,\left(V_{-\infty,\infty}(\bfy)V_{\infty, -\infty}(\bfx)\right)
\Big>\, .
\label{eq:32.5}
\ee
The second term $\left<{\cal O}\right>_2$
is obtained by changing the overall sign in
$\left<{\cal O}\right>_1$ and exchanging
$\bfx$ and $\bfy$. After some algebra~\footnote{We have used the identities
$d_{abc}t_a t_b =
{1\over 2} {N_c^2-4\over N_c} t_c$
and $d_{abc}t_a t_b t_c={(N_c^2-4)(N_c^2-1)\over 4 N_c^2} {\bf 1}$
where $\bf 1$ is the $3\times 3$ identity matrix. We have also used the cyclic
property of the trace.},
\be
\left<{\cal O}\right>_1
& = &
{g^3\,(N_c^2-4) \over 2iN_c^2 \kappa_A}
\int du^-\, \mu^6(u^-)
\int d^2 \bfu
\Bigg[{(N_c^2-1)i\over 4N_c}
\left (G^3(\bfx-\bfu) - G^3(\bfy-\bfu)\right)\, \non
 &\times& {\cal I}_1^{\infty,-\infty}
+{3i\over 2} \left( G^2(\bfy-\bfu) G(\bfx-\bfu)
- G(\bfy-\bfu) G^2(\bfx-\bfu)\right)\,{\bf
\cdot}\,{\cal I}_2^{\infty,-\infty} \Bigg] \, .
\label{eq:33}
\ee
where\, ${\cal I}_1^{\infty,-\infty}=\Big<V_{-\infty,\infty}(\bfy)V_{\infty,
-\infty}(\bfx)\Big>$ and\,\\ ${\cal I}_2^{\infty,-\infty}=\Big<t_a
V_{u^-,\infty}(\bfy)
V_{\infty, u^-}(\bfx) t_a  V_{u^-, -\infty}(\bfx) V_{-\infty,u^-}(\bfy)\Big>$.
${\cal I}_2$ can be simplified~\footnote{
The local structure in $x^-$ (see the expression after Eq.~\ref{eq:31}) ensures
that
only adjacent Wilson lines can contracted~\cite{GelisPeshier}. This simplifies
${\cal I}_2^{\infty,-\infty}$ to read
$${\cal I}_2=
t_a \left<  V_{u^-,\infty}(\bfy) V_{\infty, u^-}(\bfx) \right>
t_a \left< V_{u^-, -\infty}(\bfx) V_{-\infty,u^-}(\bfy) \right>$$.}
 using the locality of
the distributions in $x^-$. Further, since~\cite{RajLarry,GelisPeshier}
$${\cal I}_1^{v^-,u^-}
=
\exp\Big(-{g^2\,t_a t_a\over 2}
\int^{v^-}_{u^-} dz^-\, \mu^2(z^-)\int d^2\bfz\,
\left( G(\bfx-\bfz) - G(\bfy-\bfz) \right)^2
\Big)$$
and  $t_a t_a \propto {\bf 1}_{3\times 3}$, it commutes with all $t_a$, and the
two sets of
$\ave{V^\dagger V}$ in the second term of Eq.~\ref{eq:33} can be combined to
form ${\cal I}_1^{\infty,-\infty}$.  Substituting these results in
Eq.~\ref{eq:33}, and combining it with $\left<{\cal O}_2\right>$ (obtained by
 exchanging $\bfx$ and $\bfy$), we have,
\be
& &\left<{\cal O}\right>
 =
\alpha_S^3\,
{(N_c^2-4)(N_c^2-1)\over 4\,\pi\,r_0^2\,N_c^4}
A^{1/3}\,
\int {d^2\bfu}
\ln^3
{
\left| \bfx - \bfu \right|
\over
\left| \bfy - \bfu \right|
}\, ,
\non
&& {}
\times
{\rm Tr}\,
\exp\left(-{g^2\,\mu_A^2\,t_a t_a\over 2}\,\int d^2\,\bfz\,
\left( G(\bfx-\bfz) - G(\bfy-\bfz) \right)^2
\right)\, .
\label{eq:34}
\ee
We have replaced here $\int_{-\infty}^{\infty} d z^- \mu^2 (z^-) = \mu_A^2$.
Note that ${\rm Tr}\, \exp() = N_c -((N_c^2-1)/4\cdots$, so the first term in
the expansion is identical to the weak field result in Eq.~\ref{eq:26}. As
discussed
previously, the argument of the exponential here can be expressed in terms of
the saturation scale $Q_{s,A}^2$ of a
large nucleus~\cite{RajLarry,GelisPeshier,IV}. Our expression in
Eq.~\ref{eq:34} agrees with the initial condition (modulo geometrical
factors) for Odderon evolution suggested in Ref.~\cite{KSW} in a diagrammatic
approach. In the large A limit, when the saturation scale is large,
Eq.~\ref{eq:34} tells us that the Odderon contribution will be suppressed. Of
course, there is a further suppression that comes in from
quantum evolution~\cite{HIIM}.

We shall now outline the computation of the Baryon Odderon operator to all
orders in the classical color field. The full
computation, while perfectly feasible and straightforward, is extremely tedious
for reasons which will become clear.  This will be
left for a later date.
We begin with an alternative expression to that in Eq.~\ref{eq:27}, also given
in Ref.~\cite{HIIM}, which will suit us better:
\be
{\cal B}(x,y,z) &=& {1\over 3!\,2\,i}\,\left[{\rm Tr}(V_x^\dagger V_z)\,{\rm
Tr}(V_y^\dagger V_z) -{\rm Tr}(V_x^\dagger V_z V_y^\dagger V_z)
- h.c.\right]  \non &\equiv& {\cal K}_1 - {\cal K}_2 - h.c.
\label{eq:35}
\ee
To compute the expectation value of this object, we follow the same procedure
as that for the dipole Odderon operator. Consider
for example the expectation value of the first term. Using the identity in
Eq.~\ref{eq:32}, we have,
\be
\left<{\cal K}\right>_1 &=& {d_{abc}\over 12 i \kappa_A}
\int du^-\int d^2 \bfu
\Big< {\partial \over \partial \rho_a(u)}{\partial \over \partial
\rho_b(u)}{\partial \over \partial \rho_c(u)} \non
& &\left( V_{-\infty,\infty}(\bfx)
V_{\infty,-\infty}(\bfz)\right)\,\left(V_{-\infty,\infty}(\bfy)
V_{\infty,-\infty}(\bfz)\right)\Big > \, .
\label{eq:36}
\ee
One has an analogous expression for $\left<{\cal K}_2\right>$.
Unlike Eq.~\ref{eq:32.5}
where one had to manipulate $8$ terms ($2$ from each
differentiation), one now has $4^3=64$ terms from $\left<{\cal K}_1\right>$
and an equal
number from $\left<{\cal K}_2\right>$ to make a total of $128$
terms! These can be combined using Fierz identities eventually into Gaussian
correlators of 2-point, 3-point and 4-point Wilson lines.
A general technique to compute these was discussed in Appendix A of
Ref.~\cite{BGV2}.

\section{Summary}

We  showed in this note that, for SU(3) quark sources, the MV model includes a
term proportional to the cubic  Casimir in the path
integral. We obtained in Eq.~\ref{eq:17} a
quantitative expression for this correction. Remarkably, this term derived from
a random walk analysis of classical color charges is responsible for generating
Odderon excitations in the classical effective
theory of the Color Glass Condensate. Our results for the expectation values of
the dipole Odderon operator and the Baryon Odderon operator in the CGC
background field agree exactly with those of Ref.~\cite{HIIM} in the weak field
and $A=1$ limit. In addition, we computed the expectation value of the dipole
Odderon operator to all orders in the classical field and showed that the
result agreed with Ref.~\cite{KSW}.
We outlined the computation of the Baryon Odderon operator-the detailed
computation will be left for a future study.  Our results suggest
that Odderon excitations, while quantitatively suppressed, arise naturally in a
consistent mathematical framework and are therefore relevant for a systematic
study of high energy QCD.

\section*{Acknowledgments}

One of us (RV) would like to thank Jochen Bartels, Larry McLerran and Yuri
Kovchegov for very useful discussions. We would like to thank Edmond Iancu for
reading the manuscript.
RV's research is supported in part by DOE Contract No. DE-AC02-98CH10886 and by
a grant from the
Alexander von Humboldt foundation. SJ~is supported in part by the Natural
Sciences and
Engineering Research Council of Canada and by le Fonds
Nature et Technologies of Qu\'ebec.  He also
thanks RIKEN BNL Center and U.S. Department of Energy [DE-AC02-98CH10886] for
providing facilities essential for the completion of this work.


\begin{thebibliography}{99}

\bibitem{DonnachieLandshoff}A.~Donnachie and P.~V.~Landshoff,
Phys.\ Lett.\ B {\bf 296}, 227 (1992).


\bibitem{BFKL}E.~A.~Kuraev, L.~N.~Lipatov and V.~S.~Fadin,
Sov.\ Phys.\ JETP {\bf 45}, 199 (1977)
[Zh.\ Eksp.\ Teor.\ Fiz.\  {\bf 72}, 377 (1977)]; I.~I.~Balitsky and
L.~N.~Lipatov,
Sov.\ J.\ Nucl.\ Phys.\  {\bf 28}, 822 (1978)
[Yad.\ Fiz.\  {\bf 28}, 1597 (1978)].

\bibitem{FadinLipatov}V.~S.~Fadin and L.~N.~Lipatov,
  Phys.\ Lett.\ B {\bf 429}, 127 (1998); M.~Ciafaloni and G.~Camici,
  Phys.\ Lett.\ B {\bf 430}, 349 (1998).

\bibitem{reviews}A.~H.~Mueller,
arXiv:hep-ph/9911289; L.~McLerran,
Acta Phys.\ Polon.\ B {\bf 34}, 5783 (2003); N.~Armesto,
Acta Phys.\ Polon.\ B {\bf 35}, 213 (2004);
A.~M.~Stasto,
arXiv:hep-ph/0412084.

\bibitem{Nicolescu}L.~Lukaszuk and B.~Nicolescu,
  Lett.\ Nuovo Cim.\  {\bf 8}, 405 (1973).

\bibitem{BKP}J. Bartels, Nucl.\ Phys.\ B {\bf 175} (1980) 365; J. Kwiecinski
and M. Praszalowicz, Phys.\ Lett.\ B {\bf 94}, (1980) 413.

\bibitem{Odderon-solution}J. Bartels, L. N. Lipatov and G. P. Vacca, Phys.\
Lett.\ B {\bf 477}, (2000) 178; R. A. Janik and J. Wosiek,
Phys.\ Rev.\ Lett.\ {\bf 82}, (1999) 1092.

\bibitem{Carlo}C. Ewerz, hep-ph/0306137.


\bibitem{KSW}Y.~V.~Kovchegov, L.~Szymanowski and S.~Wallon,
 Phys.\ Lett.\ B {\bf 586}, 267 (2004).

\bibitem{Mueller}A.~H.~Mueller,
Nucl.\ Phys.\ B {\bf 415}, 373 (1994).

\bibitem{NZ}N.~N. Nikolaev and B.~G. Zakharov, Z. \ Phys. \ C{\bf 49}, (1991)
607.

\bibitem{BK}{I. Balitsky}, Nucl. Phys. {\bf B} {\bf 463}, 99 (1996);
{Yu.V. Kovchegov}, Phys. Rev. {\bf D} {\bf 61}, 074018 (2000).

\bibitem{Lipatov}L.~N.~Lipatov,
  Phys.\ Rept.\  {\bf 286}, 131 (1997); {\it ibid.},  {\bf 320}, 249 (1999).


\bibitem{IV}E.~Iancu and R.~Venugopalan,
arXiv:hep-ph/0303204.

\bibitem{MV}L.~D.~McLerran and R.~Venugopalan,
Phys.\ Rev.\ D {\bf 49}, 2233 (1994); {\it ibid.}, 3352, (1994); {\it ibid.},
{\bf 50}, 2225 (1994).

\bibitem{JIMWLK}J. Jalilian-Marian, A. Kovner, A. Leonidov, H. Weigert, Nucl.
Phys. {\bf B}
  {\bf 504}, 415 (1997);
{J. Jalilian-Marian, A. Kovner, A. Leonidov, H. Weigert}, Phys. Rev. {\bf D}
  {\bf 59}, 014014 (1999);
{A. Kovner, G. Milhano, H. Weigert}, Phys. Rev. {\bf D} {\bf 62}, 114005
  (2000);
{E. Iancu, A. Leonidov, L.D. McLerran}, Nucl. Phys. {\bf A} {\bf 692}, 583
  (2001);
{E. Iancu, A. Leonidov, L.D. McLerran}, Phys. Lett. {\bf B} {\bf 510}, 133
  (2001); {E. Ferreiro, E. Iancu, A. Leonidov, L.D. McLerran}, Nucl. Phys. {\bf
A} {\bf
  703}, 489 (2002).

\bibitem{HIIM}Y.~Hatta, E.~Iancu, K.~Itakura and L.~McLerran,
  arXiv:hep-ph/0501171.

\bibitem{SR}S.~Jeon and R.~Venugopalan,
Phys.\ Rev.\ D {\bf 70}, 105012 (2004).


\bibitem{JKMW}J.~Jalilian-Marian, A.~Kovner, L.~D.~McLerran and H.~Weigert,
Phys.\ Rev.\ D {\bf 55}, 5414 (1997).

\bibitem{Kovchegov}Y.~V.~Kovchegov,
Phys.\ Rev.\ D {\bf 54}, 5463 (1996); {\it ibid.}, {\bf 55}, 5445 (1997).

\bibitem{Macfarlane}
J.A.de~Azcarraga and A.J~Macfarlane,
J.\ Math.\ Phys.\  {\bf 42}, 419 (2001).

\bibitem{Collins}J.~C.~Collins,
  ``An Introduction To Renormalization, The Renormalization
  Group, And The Operator Product Expansion,''
  Cambridge Univ. Press ( 1984) 380p.

\bibitem{DoschCarlo}H.~G.~Dosch, E.~Ferreira and A.~Kramer, Phys.\ Rev.\ D {\bf
50}, (1994) 1992; H.~G.~Dosch, C.~Ewerz and V.~Schatz,
  Eur.\ Phys.\ J.\ C {\bf 24}, 561 (2002).

\bibitem{RajLarry}L.~D.~McLerran and R.~Venugopalan,
Phys.\ Rev.\ D {\bf 59}, 094002 (1999); R.~Venugopalan,
Acta Phys.\ Polon.\ B {\bf 30}, 3731 (1999).

\bibitem{GelisPeshier}F.~Gelis and A.~Peshier,
Nucl.\ Phys.\ A {\bf 697}, 879 (2002).

\bibitem{BGV2}J.~P.~Blaizot, F.~Gelis and R.~Venugopalan,
Nucl.\ Phys.\ A {\bf 743}, 57 (2004).

\end{thebibliography}
\end{document}